\documentclass{elsart5p}
\usepackage{txfonts, aas_macros}
\usepackage{graphicx, amssymb}
\journal{Astroparticle Physics}
\usepackage{subfigure}

\newcommand{\comment}[1]{}

\begin{document}
\begin{frontmatter}
\title{PoGOLite -- A High Sensitivity Balloon-Borne Soft Gamma-ray Polarimeter}

\date{Received / Accepted}

\author[SLAC]{Tuneyoshi Kamae\corauthref{cor}},
\corauth[cor]{Corresponding author.}
\ead{kamae@slac.stanford.edu}
\author[KTH,SLAC]{Viktor Andersson},
\author[TIT]{Makoto Arimoto},
\author[SU]{Magnus Axelsson},
\author[KTH]{Cecilia Marini Bettolo},
\author[SU]{Claes-Ingvar Bj\"{o}rnsson},
\author[EP]{Gilles Bogaert},
\author[KTH]{Per Carlson},
\author[SLAC]{William Craig\thanksref{now}},
\thanks[now]{Present address: Lawrence Livermore National Laboratory,
Livermore, California 94550, USA.}
\author[KTH,SLAC]{Tomas Ekeberg},
\author[KTH]{Olle Engdeg{\aa}rd},
\author[HU]{Yasushi Fukazawa},
\author[YU]{Shuichi Gunji},
\author[SU]{Linnea Hjalmarsdotter},
\author[KTH,SLAC]{Bianca Iwan},
\author[TIT]{Yoshikazu Kanai},
\author[TIT]{Jun Kataoka},
\author[TIT]{Nobuyuki Kawai},
\author[KTH]{Jaroslav Kazejev},
\author[KTH,SLAC]{M\'{o}zsi Kiss},
\author[KTH]{Wlodzimierz Klamra},
\author[KTH,SU]{Stefan Larsson},
\author[SLAC]{Grzegorz Madejski},
\author[HU]{Tsunefumi Mizuno},
\author[SLAC]{Johnny Ng},
\author[KTH]{Mark Pearce},
\author[KTH]{Felix Ryde},
\author[KTH,SLAC]{Markus Suhonen},
\author[SLAC]{Hiroyasu Tajima},
\author[HU]{Hiromitsu Takahashi},
\author[ISAS]{Tadayuki Takahashi},
\author[HU]{Takuya Tanaka},
\author[Thurston]{Timothy Thurston},
\author[TIT]{Masaru Ueno},
\author[UH]{Gary Varner},
\author[HU]{Kazuhide Yamamoto},
\author[YU]{Yuichiro Yamashita},
\author[KTH,SLAC]{Tomi Ylinen},
\author[HU]{Hiroaki Yoshida}

\address[SLAC]{Stanford Linear Accelerator Center (SLAC) and Kavli Institute
for Particle Astrophysics and Cosmology (KIPAC), Menlo Park, California 94025,
USA.}
\address[KTH]{Royal Institute of Technology, Physics Department, SE--106 91
Stockholm, Sweden.}
\address[TIT]{Tokyo Institute of Technology, Physics Department, Meguro-ku,
Tokyo 152-8550, Japan.}
\address[SU]{Stockholm University, Astronomy Department, SE--106 91 Stockholm,
Sweden.}
\address[EP]{Ecole Polytechnique, Laboratoire Leprince-Rinquet,
91128 Palaiseau Cedex, France.}
\address[HU]{Hiroshima University, Physics Department, Higashi-Hiroshima
739-8526, Japan.}
\address[YU]{Yamagata University, Physics Department, Yamagata 990-8560,
Japan.}
\address[ISAS]{JAXA, Institute of Space and Astronautical Science,
Sagamihara 229-8510, Japan.}
\address[Thurston]{The Thurston Co., 11336 30th Ave. NE, Seattle,
Washington 98125, USA.}
\address[UH]{University of Hawaii, Department of Physics and Astronomy, Honolulu,
Hawaii 96822, USA.}

\begin{abstract}
We describe a new balloon-borne instrument (PoGOLite) capable of
detecting 10\% polarisation from 200~mCrab point-like sources
between 25~keV and 80~keV in one 6~hour f\mbox{}light.
Polarisation measurements in the soft gamma-ray band are
expected to provide a powerful probe into high-energy
emission mechanisms as well as the distribution
of magnetic f\mbox{}ields, radiation
f\mbox{}ields and interstellar matter. Synchrotron radiation, inverse Compton
scattering and propagation through high magnetic f\mbox{}ields are
likely to produce high degrees of polarisation in the energy band of
the instrument. We demonstrate, through tests at accelerators, with
radioactive sources and through computer simulations,
that PoGOLite will be able to detect degrees of polarisation as
predicted by models for several classes of high-energy sources.
At present, only exploratory polarisation measurements have been
carried out in the soft gamma-ray band. Reduction of the large background
produced by cosmic-ray particles\comment{added by TK on 2007-10-12}
while securing a large effective area has been the
greatest challenge. PoGOLite uses Compton scattering and
photo-absorption in an array of 217 well-type phoswich detector cells
made of plastic and BGO scintillators surrounded by a BGO
anticoincidence shield and a thick polyethylene neutron shield.\comment{The following sentece has been changed from
``The narrow field of view (1.25~msr, 2.0~degrees $\times$ 2.0~degrees)
obtained with well-type phoswich detector
technology and the use of thick background shields
enhance the detected signal-to-noise ratio.''}
The narrow field of view
(\mbox{FWHM = 1.25~msr}, 2.0~degrees $\times$ 2.0~degrees)
obtained with detector cells and the use of thick background shields
warrant a large effective area for polarisation measurements
($\sim$228~cm$^2$ at $E=40$~keV) without sacrificing
the signal-to-noise ratio.
Simulation studies for an atmospheric overburden of
3--4~g/cm$^2$ indicate that neutrons and gamma-rays
entering the PDC assembly through the shields\comment{Was ``from the bottom and the top"}
are dominant backgrounds. Off-line event selection based on
recorded phototube waveforms and Compton kinematics
reduce the background to that expected for a $\sim$100~mCrab source
between 25~keV and 50~keV. A 6~hour observation of the Crab pulsar
will differentiate between the Polar Cap/Slot Gap, Outer Gap, and
Caustic models with greater than 5$\sigma$ signif\mbox{}icance; and also
cleanly identify the Compton ref\mbox{}lection component in the Cygnus X--1
hard state. Long-duration f\mbox{}lights will measure the dependence of the
polarisation across the cyclotron absorption line in Hercules X--1.
A scaled-down instrument will be flown as a pathfinder mission from the north of Sweden in 2010. The first science flight is planned to take place shortly thereafter.
\end{abstract}


\end{frontmatter}

\section{Introduction}
Celestial X-ray and gamma-ray sources have been
studied using their spectrum, time variability and projected image
since the early 1960s (see, for example, a review by
Fabian et al. \cite{Fabian04}).
For many sources, such observations alone do not identify the
dominant emission mechanism and polarisation measurements are
expected to add decisive information. Polarimetry will be particularly
important for studies of pulsars, accreting black holes
and jet-dominated active galaxies. Strong X-ray and gamma-ray
polarisation can arise from synchrotron emission in ordered magnetic
f\mbox{}ields, photon propagation in extremely
strong magnetic f\mbox{}ields
($>$10$^{12}$~Gauss) and anisotropic Compton scattering,
as has been discussed in Siddons~\cite{ReviewPol91} and
reviewed by Lei et al.~\cite{Lei97}. The orientation of the polarisation
plane probes the intensity and direction of the magnetic
and radiation f\mbox{}ields, as well as
the matter distribution around sources.

Despite the potential importance of polarisation measurements, the Crab nebula
is the only source outside the solar system from which polarisation has been signif\mbox{}icantly detected in this energy range. The f\mbox{}irst clear detection of polarisation was at 2.6~keV and 5.2~keV, with an instrument on board the OSO--8 satellite (Weisskopf et al. \cite{Weisskopf76,Weisskopf78}). Due to the limited effective area, the instrument could not detect
polarisation from the Crab pulsar or other sources
(Silver et al. \cite{Silver78}; Long et al. \cite{Long80}).

A polarisation detection has been reported from a gamma-ray burst observed with an instrument on the RHESSI satellite in 2002 (Coburn \& Boggs \cite{Coburn03}).
This claim has been seriously challenged and hence remains
controversial (Rutledge \& Fox \cite{Rutledge04}; Wigger et al.~\cite{Wigger04}).
A new detection of polarisation in two gamma-ray bursts archived in the BATSE catalog has been published recently (Willis et al. \cite{Willis05}).
The authors f\mbox{}itted the observed time-binned counting rates
including polarisation-dependent Compton ref\mbox{}lection
off the atmosphere.
The f\mbox{}it indicates a polarisation degree exceeding 35--50\%.

Two techniques have been developed
to measure hard X-ray (soft gamma-ray) polarisation from
astrophysical sources.
Below 10 keV, polarisation can be deduced by tracking the electron
from photo-absorption in an imaging gas detector. From about 25~keV to 1~MeV, polarisation can be determined by measuring the azimuthal angle
distribution of Compton scattered photons as has been
exemplif\mbox{}ied in papers contained in the proceedings edited by
Turner \& Hasinger~\cite{Turner06}.
Due to atmospheric opacity, instruments have to be
launched into a satellite orbit if the first technique is to be used.
Instruments relying on the latter technique can be f\mbox{}lown
on high altitude balloons since photons with energies greater
than $20-25$~keV
reach the instrument through typical
atmospheric overburdens of 3--4~g/cm$^2$.
In this energy band, however, the large background produced by cosmic-ray
particles poses significant detector design challenges.
Several instruments are under development to detect X-ray or gamma-ray
polarisation from astronomical sources using the Compton scattering technique
in balloon and satellite experiments. An important figure of merit
for these instruments is the lowest degree of polarisation
detectable at the 3$\sigma$ level, referred to as the minimum detectable polarisation. Most instruments have gamma-ray bursts as their primary target. Examples include
POLAR (Produit et al. \cite{Produit05}, 10--300~keV, satellite) and
GRAPE (Legere et al. \cite{Legere05},
50--300~keV, long-duration balloon or satellite).
PHENEX (Gunji et al. \cite{Gunji03})
is designed to detect a 10\% polarisation from the Crab nebula in the
40--300~keV range in a 3~hour balloon observation.
CIPHER (Silva et al. \cite{Silva03}) will study steady-state objects between 10~keV and
1~MeV with the capability of detecting 5\% polarisation from the Crab nebula
in a 3~hour satellite orbit observation.

The Polarised  Gamma-ray Observer -- Light-weight version (PoGOLite)
is a balloon-borne astronomical soft gamma-ray polarimeter optimised for
point-like sources. It measures
polarisation in the energy range 25--80~keV from sources as low as 200~mCrab by using the azimuthal
angle anisotropy of Compton-scattered photons.\comment{The followint sentence has been added by TK on 2007-10-12. Modified on 2007-10-19 by Mark Pearce} Several features distinguish PoGOLite from other balloon-borne instruments
proposed to observe astronomical sources using the Compton scattering
technique such as PHENEX \cite{Gunji03, Turner06}:
\begin{itemize}
\item The narrowest field of view: $1.25$~msr (FWHM).
\item The largest effective area between 25~keV and 60~keV:
$\sim$228~cm$^2$ for polarisation measurements at 40~keV.
\item Sensitivity extends as low as to 25~keV to address the sources and processes mentioned above.
\item The lowest background rate: $\sim$100~mCrab for 25--50~keV.
\item A slightly lower modulation factor, $\sim$33\%.
\end{itemize}

The instrument is currently under construction and an engineering flight of a 61 unit ``pathfinder'' instrument is planned for
2010 from the Esrange facility in the North of Sweden.
The instrument was originally designed to record Compton scattering
and photo-absorption in an array of 397 phoswich detector cells made of
plastic and BGO scintillators, surrounded
by active BGO anticoincidence shields (called PoGO).
Through trade-off studies including detector simulation and design, cost estimation and prototype testing, we converged on a lighter version of the original
PoGO instrument (Andersson et al. \cite{Andersson04}).
This new design (PoGOLite) will be able to reach
a higher altitude (41--42~km with a 1 million m$^3$ balloon) and extend the
lower energy limit down to 25~keV. The lighter design also simplif\mbox{}ies
implementation of the mechanism which permits the polarimeter to rotate around its longitudinal axis,
which is essential in reducing systematic errors in polarisation measurements.
The pointing system and gondola design are inspired
by the f\mbox{}light-proven design of High Energy Focusing Telescope (HEFT,
Gunderson et al. \cite{Gunderson04})
and Balloon-borne Large-Aperture Sub-millimeter Telescope
(BLAST, Pascale et al. \cite{Pascale08}).
The overall design of the PoGOLite polarimeter and gondola has
incorporated features needed to accomplish long-duration balloon f\mbox{}lights
from Sweden to North America in the future.

In this paper we describe the key PoGOLite polarimeter design features,
summarise the results obtained during tests with prototype
instruments, and present the scientific potential.
Relevant theoretical models for high-energy emission
from selected targets will be discussed together with simulated measurements
thereof for one 6 hour balloon f\mbox{}light. We show that PoGOLite will
open a new obervational window on high-energy astrophysics,
with the promise of clarifying the emission mechanism of many sources.


\section{PoGOLite Detector and Data Processing}

\subsection{PoGOLite Detector}
\begin{figure*}[th] 
\centering
\includegraphics[width=14cm]{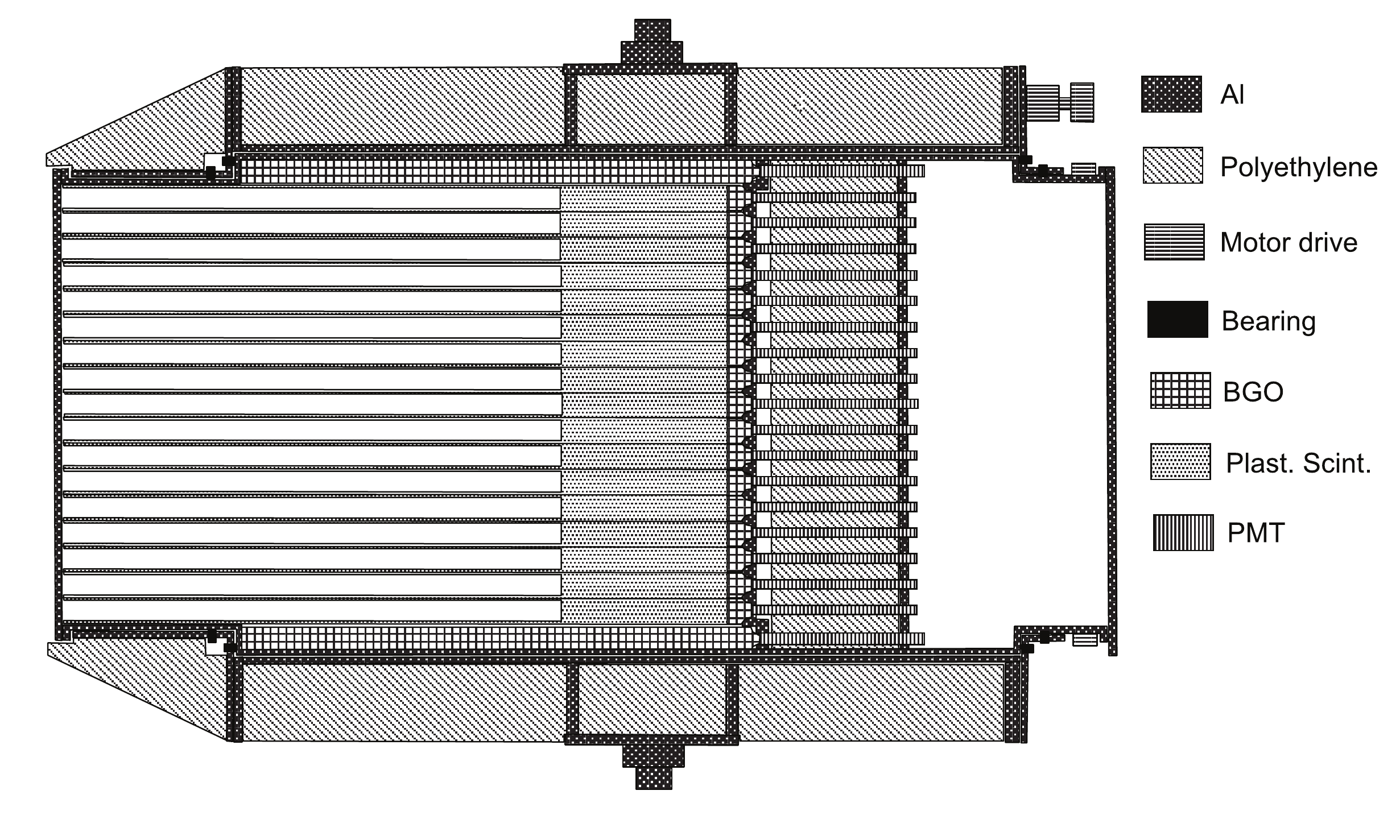}
\caption{Axial cross-section of the PoGOLite polarimeter.
The inner cylinder holds 54 BGO crystal modules, which comprise the Side Anticoincidence Shield (SAS), 217 Phoswich Detector Cells (PDCs), 271 photomultiplier
tubes (PMTs), a data acquisition system and the lower polyethylene
neutron shield. The outer cylinder houses the mechanism which allows
the polarimeter to rotate around the
longitudinal axis, the pivot for elevation pointing,
and the lateral polyethylene neutron shield.
The assembly is $\sim$140~cm in length and
$\sim$100~cm in diameter and is estimated to weigh about 800~kg.}
\label{PoGOLitePolBW}
\end{figure*}

\begin{figure}[th] 
\centering
\includegraphics[width=8.8cm]{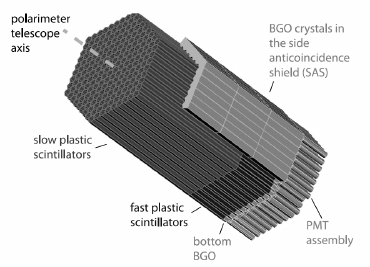}
\caption{\label{PoGO_3D}An overview of the detector arrangement in the PoGOLite polarimeter. The outermost units are the SAS BGO modules. For clarity, not all modules are shown, so the 217 PDCs are partially exposed, showing the slow and
fast scintillators, the bottom BGO crystals, and the PMTs.
The mechanical structures and the neutron shield are not shown.}
\end{figure}

\begin{figure} 
\centering
\includegraphics[width=8.8cm]{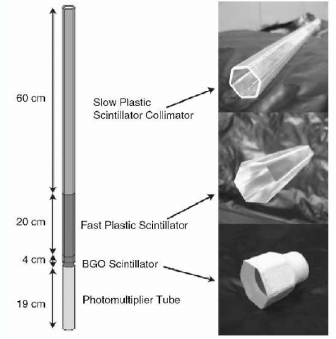}
\caption{One Phoswich Detector Cell: (from top to bottom)
60~cm long slow plastic scintillator well, 20~cm long fast scintillator
rod, 4~cm long BGO crystal and 19~cm long phototube assembly. All
scintillators are covered with ref\mbox{}lective materials, VM2000 or BaSO$_4$.
In addition, the well-portion of the PDC is wrapped
with 50~$\mu$m thick lead and tin foils for passive collimation.}\label{PoGOLitePDCBW}
\end{figure}

The PoGOLite detector consists of a hexagonal close-packed array of 217
well-type phoswich detector cells (PDCs) and 54 side anticoincidence shield
(SAS) detectors
made of bismuth germanate oxide (BGO) scintillators. This detector assembly is housed in a rotating cylindrical structure (the inner cylinder), which is placed inside another cylindrical structure reinforced with ribs and f\mbox{}illed with polyethylene (the outer cylinder) as shown in Fig.~\ref{PoGOLitePolBW}.
The arrangement of the PDCs is detailed in Fig.~\ref{PoGO_3D}.
Each PDC, Fig.~\ref{PoGOLitePDCBW},
is composed
of a thin-walled tube (well) of slow plastic scintillator at the top
(Eljen Technology EJ--240, f\mbox{}luorescence decay time 285~ns),
a solid rod of fast plastic scintillator
(Eljen Technology EJ--204, decay time $\sim$2~ns), and a bottom BGO crystal (Nikolaev Institute of Inorganic Chemistry,
decay time $\sim$300~ns), all viewed by
one photomultiplier tube (Hamamatsu Photonics R7899EGKNP).
The thin-walled well is wrapped with a reflective layer (3M VM2000) and
50~$\mu$m thick foils of lead and tin, and the bottom BGO is coated with
a reflective BaSO$_4$ layer. In the PDCs, the wells serve as active collimators,
the fast scintillator rods as active photon detectors, and the bottom
BGOs as a lower active shield. The overall length of the active collimator, 600~mm, together with the
24~mm mean diameter well opening, sets the
solid angle event acceptance, which is 1.25 msr
(2.0~degrees $\times$ 2.0~degrees).
Signals from each PMT are continuously sampled, digitised and
recorded at 36~MHz.
Through on board examination of the recorded waveforms, signals consistent with
being fast scintillator light are selected from those mixed
with slow/BGO scintillator light.
This well-type phoswich detector technology was developed and used
to reduce the cosmic ray-induced backgrounds by more than one order
of magnitude for the WELCOME balloon experiments
(Kamae et al. \cite{Kamae92,Kamae93};
Takahashi et al. \cite{Takahashi92,Takahashi93}; Gunji et al.
\cite{Gunji92,Gunji94}; Miyazaki et al. \cite{Miyazaki96};
Yamasaki et al. \cite{Yamasaki97}). Based on this success,
the technology has been applied to the Suzaku Hard X-ray Detector
(Kamae et al. \cite{Kamae96}; Makishima et al. \cite{Makishima01};
Kokubun et al. \cite{Kokubun04}). The Hard X-ray Detector is operating
in orbit with the lowest background achieved in its energy band, 12--600~keV
(Takahashi et al. \cite{Takahashi07}; Kokubun et al. \cite{Kokubun07}).

\begin{figure}[htp] 
\centering
\includegraphics[width=8.8cm]{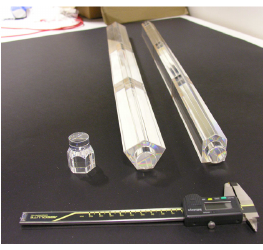}
\caption{The three types of BGO modules and crystals used in
the PoGOLite anticoincidence system.
A corner module (center) and an edge module (right) of the SAS are
shown, together with a bottom BGO crystal of the PDC (left).}
\label{PoGOLiteSASBW}
\end{figure}

The side anticoincidence system (Marini Bettolo et al. \cite{cmb})
consists of 54 modules of BGO crystals which cover two thirds of
the height of the PDC elements. Each module is built from
three crystals glued together, as shown in Fig.~\ref{PoGOLiteSASBW}.
The crystals have a pentagonal cross-section and are tightly packed
around the PDC assembly. A module-to-module gap of $\sim$100 $\mu$m
is foreseen, mainly due to the
reflective BaSO$_4$ layer applied to each crystal.
The BGO crystals are supplied by the Nikolaev Institute of Inorganic
Chemistry and are read out with the same type of phototube that is used
for the PDC units.

With an estimated total side anticoincidence rate
of about 100~kHz at float altitude, a simple veto would reject
about 6\% of detected events by random coincidence.
The segmentation allows the side anticounter and the PDC hits to be correlated,
which will further reduce the number of valid events rejected.
Segmentation also allows possible
asymmetries in the backgrounds to be studied and corrections
to be applied in off-line analysis. The anticoincidence
threshold is around 75~keV.

\comment{The following sentences were added by TK on Jan 24 2008}
In the PoGOLite design, the PDCs serve to detect and measure
both Compton scattering(s) and photo-absorption of astronomical
soft gamma-rays. Hence the design is scalable in size and
facilitates end-to-end tests from developmental stage.
If the two functions had been assigned to different
components, it would not have been easy to attain
a large effective area nor to perform end-to-end tests.

\subsection{Data Processing}

\begin{figure*}
\centering
\includegraphics[width=18cm]{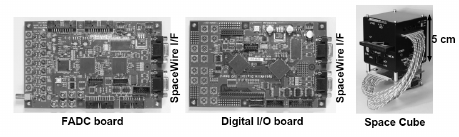}
\caption{The flight model Flash ADC board, Digital I/O board,
and Space Cube.}
\label{DAQ1}
\end{figure*}

Signals from all 217 PDCs and 54 SAS PMTs are fed to individual
f\mbox{}lash ADCs on 38 Flash ADC (FADC) Boards (see Fig.~\ref{DAQ1})
and digitised to 12 bit accuracy at 36~MHz.
Field programmable gate arrays (FPGAs) check for a transient signal
compatible with an energy deposition in the fast scintillator
("clean fast signal") between a minimum ($\sim$15~keV) and
a maximum ($\sim$200~keV).
A veto signal is issued when the FPGAs detect a transient signal
exceeding the upper discrimination level to suppress cosmic ray backgrounds
or signals compatible with the rise time characteristic of the slow scintillator or the bottom BGO.
Digital I/O (DIO) boards (see Fig.~\ref{DAQ1}) collect the trigger signals from eight FADC boards and process local trigger signals (both clean fast signals and veto signals).
A global DIO board collects local trigger signals from DIO boards and processes a global trigger signal.
If a clean fast signal is found without a veto signal
in a time-window, a global trigger is issued
and digitised waveforms are stored for a period of
$\sim$1.6~$\mu$s starting $\sim$0.4~$\mu$s earlier than the trigger.
Only PDC data with a transient signal greater than a value corresponding to around
0.3 photo-electron (0.5~keV) will be stored in the buffer of
the FADC boards for data acquisition. Note that we conservatively assume the threshold to be
1.0 photo-electron in all computer simulations.
\comment{HT: we have not decided real threshold value for online zero suppression yet. It depends on the data rate we can tolerate.}
The FADC boards also save information on what channels have stored waveforms for all triggered events (we call it the hit pattern).
In order to minimize the dead time, the DIO boards start accepting triggers as soon as it has received signals from all FADC boards that all waveforms are buffered.
The trigger rate is expected to be about 0.5~kHz
resulting in a dead time fraction of about 0.6\%
and a data-rate of about 240~kB/s or 0.9~GB/hr with zero suppression.
An onboard computer, the Space Cube (see Fig.~\ref{DAQ1}), collects hit patterns and waveforms tagged by trigger identification numbers from FADC boards asynchronously for dead-time-less data acquisition and records them into f\mbox{}lash memory drives.
The Space Cube also collects the hit pattern from FADC boards to confirm that all waveforms are collected for every event.

\begin{figure}[htp]
\centering
\includegraphics[width=8.8cm]{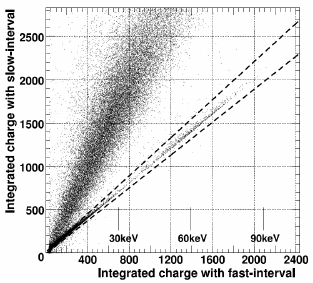}
\caption{Selection of clean gamma-ray hits from $^{241}$Am
on the fast scintillator rod while the slow scintillator well is irradiated with electrons from $^{90}$Sr.
Each point corresponds to one triggered event: the charge collected
in a short interval
(120~ns) gives the abscissa and that in a long interval
(1~$\mu$s) the ordinate.  Signals in the fast scintillator
rod are gamma-rays from $^{241}$Am (a line at 60~keV and several lines
between 15 and 36~keV) lying between the two dashed lines.
Background electron signals form the thick band to the
left of the dashed lines. A crude energy scale for gamma-rays
detected in the fast scintillator has been added.
}\label{Sr90psd2}
\end{figure}

The recorded data is processed in the following steps:

\noindent {\bf Step 1:} Select the PDC where a photo-absorption took
place by choosing the highest energy deposition compatible with being a clean fast signal
in the fast scintillator. Fig.~\ref{Sr90psd2} shows how clean fast
signals in the fast scintillator of gamma-rays from $^{241}$Am
are selected in a high background environment created by electrons from $^{90}$Sr irradiating the slow scintillator. Each dot in the f\mbox{}igure represents one recorded event: the horizontal and vertical coordinates are the charges integrated
over the fast ($\sim$120~ns) and slow ($\sim$1~$\mu$s) intervals
respectively. The diagonal
concentration of dots between the two dashed lines corresponds to
clean fast signals in the fast scintillator.

Data in Fig.~\ref{Sr90psd2}
indicate strong
concentrations around the dominant line at 60~keV and the
several weaker lines between 15 and 36~keV
in the diagonal slice, all produced by $^{241}$Am gamma-rays
in the fast scintilltor.
Included in the lower-energy concentration are Compton scattered
events where only a fraction of the gamma-ray energy is deposited in the
fast scintillator. The site with the highest clean fast signal pulse-height
is selected as the photo-absorption site.

\noindent {\bf Step 2}: Waveforms from all neighboring PDCs
(up to two layers or 18 PDCs) around the identified photo-absorption
site are searched for a Compton scattering signal.
\comment{The following 2 sentences were added by TK on 2007-10-12. There was an error and corrected on 2007-10-19.}
We estimate the chance of finding 1 \comment{Added the following 2 words}
Compton scattering hit to be about 20.5\%,
2 hits about 10.5\% and more than 3 hits about 15.5\%
when averaged between 20 and 80~keV. We plan to accept the events with
1 and 2 Compton scattering sites.
In the beam tests, the Compton site threshold was set to
$\sim 0.3$ photoelectrons where 0.65 photoelectron corresponds to
1~keV in the fast scintillator.  If more than one
Compton site is found, the site with highest pulse-height is chosen.
This strategy is justified by the fact that a low energy Compton recoil electron means that there is less effect on the azimuthal scattering angle.
Our computer simulation shows that the modulation factor of two Compton site events is $\sim$23\% while it is $\sim$35\% for one Compton site event.
We conservatively
assumed a 1.0 photoelectron threshold for detecting Compton scattering
in all simulations.

\begin{figure}[hbtp] 
\centering
\includegraphics[width=8.8cm]{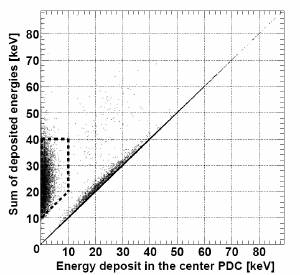}
\caption{Selection of Compton-scattered coincidence events
for data obtained with a 25~keV polarised beam at the KEK Photon Factory.
The beam is arranged to scatter in the central PDC and be photo-absorbed in a
surrounding PDC (cf. Fig.~\ref{GeometryPrototype}). The dashed box shows the valid kinematic
area for Compton scattering events where a small amplitude
fast signal from the recoil electron and a larger amplitude
fast signal from the photoabsorption are registered.
}\label{KinematicCut25keV}
\end{figure}

\begin{figure} 
\centering
\includegraphics[width=5cm]{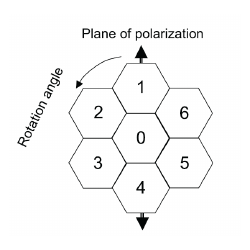}
\caption{Arrangement of the 7 PDCs in the KEK 2007 beam test.
The polarised 25~keV pencil beam hits the center of PDC~0.
Photo-absorption was recorded in one of the 6 peripheral PDCs.
The set-up was rotated in azimuth in 15~degree steps.}
\label{GeometryPrototype}
\end{figure}

\noindent {\bf Step 3}: Distributions of the sum of the two energy depositions (the photo-absorption site and the Compton scattering site) versus the energy deposition at the Compton scattering site are studied for the selection of valid events. Fig.~\ref{KinematicCut25keV} shows such a distribution obtained with a polarised photon beam at the KEK Photon Factory, Japan, in 2007.
Seven flight-model PDCs were arranged as shown in Fig.~\ref{GeometryPrototype}. The area enclosed by the three dashed-line segments in Fig.~\ref{KinematicCut25keV} denotes the allowed kinematical region for Compton scattered and photo-absorbed events.
For these selected events, we define the azimuthal angle of scattering
by a line connecting the centers of the two PDCs where
Compton scattering and photo-absorption are detected.  Since the
polarimeter orientation drifts constantly in the celestial coordinate
system during observations and since the polarimeter will be rotated
around its axis, the azimuth angle fixed thereon has to be aligned
against the position angle around the target object.  The distribution of
scattering position angles aligned in this way shows a sinusoidal
modulation when the incident gamma-rays are polarised.
This measured distribution
is fitted with a sinusoidal curve with a constant offset, the modulation curve.
The modulation factor is defined as
the ratio between the amplitude of the sinusoidal part and the offset.

\begin{figure} 
\centering
\includegraphics[width=9cm]{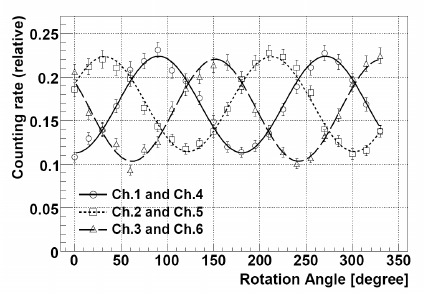}
\caption{Modulation measured in a polarised 25~keV pencil beam
at the KEK Photon Factory. The beam polarisation was ($90\pm1$)\% and the measured
modulation factor is ($33.8\pm 0.7$)\%.}\label{Modulation25keV}
\end{figure}

Fig.~\ref{Modulation25keV}
shows three sets of modulation curves
obtained for three coplanar pairs of PDCs in the KEK 2007 prototype.
The abscissa is the rotation angle of the prototype
relative to the plane of the beam polarisation.
The PDC arrangement is shown in Fig.~\ref{GeometryPrototype}.
The beam had an energy and polarisation of 25~keV and ($90\pm1$)\%, respectively. It was aimed at the center of PDC~0 and the 6 possible
coincidence pairs (PDC~0--PDC~1, PDC~0--PDC~2, PDC~0--PDC~3, PDC~0--PDC~4,
PDC~0--PDC~5 and PDC~0--PDC~6) were combined into three coplanar sets
in the f\mbox{}igure. The azimuthal angle of Compton scattering
is offset by 0, 60 and 120~degrees for the three coplanar pairs
PDC~1--PDC~4, PDC~2--PDC~5 and PDC~3--PDC~6, respectively. The measured
modulation factor is ($33.8\pm 0.7$)\%, in line with expectations.
Note that our computer simulation predicts a lower modulation
factor when gamma-rays are distributed over the entire cross-section
of the PDC, as tabulated in Table~\ref{Table1}.
\begin{table*}
\centering
\caption{Expected performance characteristics of PoGOLite at an atmospheric overburden of 4~g/cm$^2$}
\label{Performance}
\begin{tabular}{l c c c c c c c}
\hline\hline
 & 25~keV & 30~keV & 40~keV & 50~keV & 60~keV & 80~keV & Total/Average \\
\hline
Min Detectable Pol. in 6 hrs & \multicolumn{7}{c}{} \\
for a 100~mCrab source & \multicolumn{5}{c}{10.5\%} \\
for a 200~mCrab source & \multicolumn{5}{c}{6.5\%} \\
\hline
Field of view & \multicolumn{7}{c}{1.25~msr (2.0~degrees $\times$ 2.0~degrees)} \\
\hline
Time resolution & \multicolumn{7}{c}{1.0~$\mu$s} \\
\hline
Geometric area & \multicolumn{7}{c}{994~cm$^2$} \\
\hline
Eff. area for pol. measurement & 93~cm$^2$ & 167~cm$^2$ & 228~cm$^2$
               & 198~cm$^2$ & 172~cm$^2$ & 158~cm$^2$ & \\
(Attenuation by air not included) & & & & & & & \\
\hline
Signal rate for a 100~mCrab source & 0.039~/s/keV & 0.044~/s/keV & 0.039~/s/keV                & 0.025~/s/keV & 0.015~/s/keV & 0.0056~/s/keV & 1.52~/s \\
\hline
Background rate & 0.043~/s/keV & 0.041~/s/keV & 0.030~/s/keV & 0.029~/s/keV
               & 0.022~/s/keV & 0.018~/s/keV & 1.77~/s \\
\hline
Modulation for 100~\% polarised & 33\% & 29\% & 26\% & 27\%
                               & 32\% & 40\% & 32.5\% \\
beam with Crab spectrum & & & & & & & \\
\hline
\end{tabular}\label{Table1}
\end{table*}
\comment{The following 2 sentences added by TK on 2007-10-12 and
modified by Mark Pearce on 2007-10-19. In Jan 2008 the two senteces
have been removed because they are in Table 1:
``From simulations, the modulation factor at 25~keV is found
to be $\sim 35$\% for events with a single Compton scattering site,
and $\sim 23$\% for those with double Compton scattering sites.
When averaged over the Crab nebula
spectrum and the single and double Compton scattering events,
the modulation factor is $\sim 32$\%."}

\subsection{Background Suppression}
Potential backgrounds affecting the polarisation measurement can arise
from extraneous gamma-ray sources within the f\mbox{}ield-of-view,
gamma-rays and charged particles that leak through the side and bottom BGO
anticoincidence systems, and neutrons produced in the atmosphere
and the gondola structure. For high latitude flights, auroral X-ray
emission due to bremsstrahlung emitted at altitudes of $\sim$100~km is
also a potential background.

The bottom BGO crystals integrated into the 217 PDCs, in combination with the 54 BGO crystal assemblies in the side anticoincidence system (SAS), efficiently suppress
the gamma-ray background together with
the slow plastic scintillator tubes and the lead/tin foils around the tubes.
Polyethylene walls surround the aluminum structures.
The averaged thickness of the
polyethylene is about 10~cm in the upper part and about 15~cm
in the lower portion, as shown in Fig.~\ref{PoGOLitePolBW}.
These walls, along with the slow plastic scintillator tubes, slow
down fast (MeV) neutrons to keV energies, thereby making it less likely that
a trigger is generated and the selection criteria
described in the previous section satisfied.

\begin{figure}[hbtp] 
\centering
\includegraphics[width=8cm]{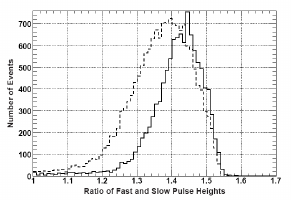}
\caption{Distributions of the ratio of the pulse heights
integrated with the fast (120~ns) and slow (1~$\mu$s) time constants
for neutron-induced (dashed) and gamma-ray (solid)
events in the energy band between 45 and 75~keV.
The neutrons are from $^{252}$Cf and the gamma-rays are from $^{241}$Am.
About 51\% of neutron-induced background events can be rejected
while keeping about 72\% of the signals.}\label{NeutronReject1}
\end{figure}

The PoGOLite design adopts five new background suppression
schemes compared to traditional phoswich technology: (1) an active collimator
limits the f\mbox{}ield of view to 1.25~msr (2.0~degrees $\times$ 2.0~degrees); (2) a thick polyethylene neutron
shield is included; (3) all non-zero PDC waveforms are sampled at 36~MHz for
about 1.6~$\mu$s starting 0.4~$\mu$s before the trigger to eliminate fake
events produced by pulse pile-up;
(4) additional suppression of neutron-induced background is possible using
recorded PDC waveforms as shown in Fig.~\ref{NeutronReject1}; (5) the counting rate and energy spectrum are recorded for all 54 BGO assemblies of the SAS to facilitate correction for possible azimuthal modulation in background events, as discussed further below.

\subsubsection{Background Simulation}
\label{Background sim}
Background rates have been calculated using simulation programs
based on the Geant4
package (Brun et al. \cite{Brun87}; Amako et al. \cite{Amako94}).
The program incorporates background flux models,
the PoGOLite detector geometry, the trigger logic,\comment{Added by TK} energy-loss due to scintillation light conversion,
and the on board and off-line event f\mbox{}iltering algorithm.

Background gamma-rays can reach the fast scintillators by crossing the
BGO crystals unconverted as well as through gaps between the BGO crystals.
The gamma-ray background f\mbox{}lux model has been developed
based on observational data (Mizuno et al. \cite{Mizuno04}) and generates
diffuse cosmic and atmospheric gamma-rays
representative of a balloon environment
with 4~g/cm$^2$ atmospheric overburden.

Neutrons can fake valid Compton scattering events through elastic
scattering off protons in the fast scintillators. Two or more clean
fast signals have to be produced\comment{A few words removed in Jan 2008 per Mizuno san's comment:
``within a distance corresponding to 2 PDC widths and"}
with kinematics compatible with Compton scattering. Since the
neutron mean-free-path in the plastic scintillator is
less than 1~cm for kinetic energies below 3~MeV, fake
events cannot come from neutrons with energies less than $\sim$1~MeV.
Neutrons must also escape detection in the BGO scintillator.
In this regard, we note that the gamma-ray emitting inelastic
cross-sections exceed 1 barn in bismuth for neutrons with energies greater than $\sim$5~MeV. A combination of these two conditions limits
background neutrons to the \comment{Was 1-10MeV but changed per Mizuno's
comment} 1--100~MeV range.
The scintillation light yield in one fast
scintillator must be consistent with that for the photo-absorption
site in the PoGOLite energy band, and the other one with that for the
Compton scattering site.  Scintillation light emitted by the
low-energy recoil proton in n-p elastic scattering is suppressed as
described by Verbinski et al.~\cite{Verbinski68} and
Uwamino et al.~\cite{Uwamino82}.
Our simulation program estimates these processes
with dedicated programs within the Geant4 framework using
the QGSP\_BERT\_HP physics
list (Koi et al. \cite{Koi07}).  We note that 96\% of the neutron
f\mbox{}lux is attenuated by the combination of a 10~cm thick
polyethylene wall and 6~cm slow scintillator array.
More details can be found in Kazejev~\cite{Kazejev07}.

The neutron f\mbox{}lux used in the simulation is from
a model proposed by Armstrong et al.~\cite{Armstrong73} for an
atmospheric depth of 5~g/cm$^2$. The angular distribution is assumed
to be isotropic, but we note that according to
Armstrong et al.~\cite{Armstrong97}, the upward
and downward f\mbox{}luxes are typically 80\% and 20\%
of the isotropic f\mbox{}lux, respectively.
This anistropy in background neutron distribution therefore affects the
background rate at less than $\sim$20\% level.

The neutron f\mbox{}lux produced by cosmic rays in the passive
components of the payload has been calculated to be negligible
based upon an empirical formula developed by Cugnon et al.~\cite{Cugnon97}.

Background due to charged particle interactions has been determined
to be negligible. This background is eliminated by requiring that
the pulse-height lies between lower and upper discrimination levels,
and by the anticoincidence function of the BGO shield embedded
in the PDC assembly and the SAS.

The background presented by auroral X-rays in high latitude flights
(e.g. from the Esrange facility in Northern Sweden) has been studied
by Larsson et al.~\cite{Larsson07}. In a conservative scenario, significant
auroral X-ray emission is predicted for less than 10\% of the flight
time during a long duration flight from Esrange. The resulting X-ray
fluxes in the PoGOLite energy range are expected to be of a few tens
of mCrab. Short duration spikes exceeding 100 mCrab are also possible.
In order to identify periods with significant auroral activity,
PoGOLite will be equipped with auroral monitoring instruments
(narrow passband photometers and magnetometers). It is noted that
although a background for PoGOLite, polarization measurements of
auroral emission will provide unique results, allowing the pitch angle
distribution of the incident electrons to be determined, thereby
giving essential clues to the acceleration processes at play.

We also expect higher charged particle,
neutron and gamma-ray backgrounds during flights from Northern Sweden
than from Texas or New Mexico, USA.
We plan to assess backgrounds during an engineering flight from Sweden and increase the thickness
of the neutron shield before making long-duration flights if needed.

\begin{figure}[hbtp] 
\centering
\includegraphics[width=9cm]{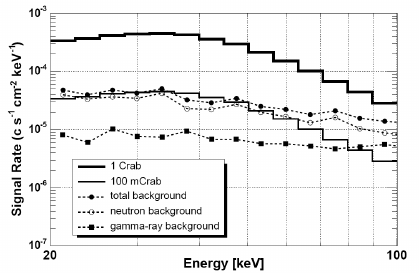}
\caption{Expected background rates at an atmospheric overburden
of 4~g/cm$^2$ compared with signal rates expected
for a Crab source (thick solid histogram) and
a 100~mCrab source (thin solid histogram). Filled circles denote
the total background, open circles the neutron background and
filled squares the gamma-ray background.}
\label{BkgdMay07BW}
\end{figure}

The total background rate is shown in Fig.~\ref{BkgdMay07BW} with
filled circles, the neutron background rate with open circles, and
the gamma background rate with filled squares. These rates can be
compared with signal coincidence rates for a 1 Crab source
(thick solid histogram) and a 100 mCrab source (thin solid histogram).
The expected total background coincidence rate is equivalent to
that of a $\sim$100~mCrab source
between 25 and 50~keV. Our simulation program predicts about 60--70\% of the total background to be produced by albedo neutrons and the rest by albedo gamma-rays. The contribution from charged cosmic rays will be less than 10\%.

\subsubsection{Minimizing Possible Systematic Bias}

Backgrounds can affect polarisation measurements in two different ways.
First, the minimum detectable polarisation deteriorates due to
statistical f\mbox{}luctuations in the background counts.
For most polarised
astronomical sources, 10--20\% polarisation is expected. This means
an instrument must be able to detect a 3--6\% modulation factor.
Instrumental response must be axially symmetric to better than
a few percent.
Various sources of asymmetry are expected: anisotropy in background
events, systematic offset in the pointing and asymmetry in the
instrument response.

To reduce systematic bias, the
PDC assembly will be rotated axially in 15 degree steps.
Such a rotation mechanism has been implemented in the beam tests
and the measured modulation factors for six combinations of
photo-absorption and Compton scattering sites have agreed
to within $\sim$2\% in absolute value.

Second, the background due to albedo neutrons will be anisotropic
when the polarimeter axis is tilted off the zenith. Such an anisotropy
will be monitored by recording counting rates for all
54 BGO crystal assemblies of the SAS and for the outermost 48
PDC units. The measured anistropy will be used to remove modulation artifacts
introduced by background events.

\section{Performance Verification with Accelerator Beams and
Computer Simulations}

Various prototype PDCs have been tested in polarised gamma-ray beams
to verify the simulation program -- including physics processes
implemented in Geant4 -- and subsequently to guide optimisation of
trigger thresholds and the event selection algorithm. In 2003, a
beam test was conducted at the Argonne National Laboratory, in which the
analysis on the azimuthal modulation demonstrated errors in
the treatment of photon polarisation in the Compton/Rayleigh scattering
processes implemented in
Geant4 (Mizuno et al. \cite{Mizuno05}). The next test, conducted at
KEK in Japan, was aimed at studying the low energy response of the PMT assemblies
(Kataoka et al. \cite{Kataoka05}).
One f\mbox{}light PDC and 6 prototype PDCs were then tested at KEK\comment{added two words} in 2004
and the revised Geant4 model was verif\mbox{}ied to $\sim$3\%
(absolute) in modulation factor (Kanai et al. \cite{Kanai07}).
The most recent test at KEK was completed in
March 2007 with 7 f\mbox{}light-version PDCs and front-end
electronics.  The measured modulation for ($90\pm1$)\% polarised 25~keV
gamma-rays is presented in Fig.~\ref{Modulation25keV} and the details
will be described in a separate publication (Takahashi et
al. \cite{takahashi08}).

\begin{figure}[hbtp] 
\centering
\includegraphics[width=.25\textwidth]{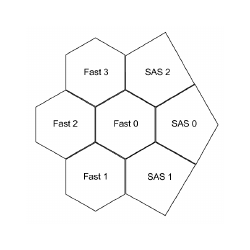}
\caption{\label{Gothenburg array}The detector geometry used in the neutron test. The width of the array is about \mbox{10 cm}. A polyethylene shield with a thickness of \mbox{10 cm} (not shown here) surrounds the entire detector array. The SAS BGO shields are facing the neutron generator (not shown in this picture) to resemble the situation with atmospheric neutrons incident on the side anticoincidence shield surrounding the instrument.}
\end{figure}

\begin{figure}[hbtp] 
\centering
\includegraphics[width=.475\textwidth]{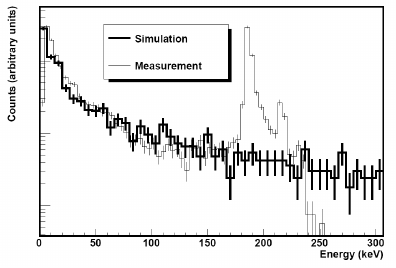}
\caption{\label{SimMeas_10cm}Neutron spectra from simulation (thick solid line) and from measurements (thin solid line) for a plastic scintillator in a simple detector array surrounded by a \mbox{10 cm} thick polyethylene shield. The two peaks in the measured spectrum are from saturation in the electronics and are therefore not expected in the simulation. The distributions are normalized to the value at \mbox{10 keV}.}
\end{figure}

The response of the PDC fast plastic scintillator to neutrons in the
MeV energy range has been studied in great detail. Tests with neutrons produced
in the decay of $^{252}$Cf confirm that the neutron-induced background
can be reduced by about 49\% while sacrificing about 28\% of gamma-ray
events between 45~keV and 75~keV (see Fig.~\ref{NeutronReject1}). This
filtering can be applied during off-line analysis to the outer-most
layer of PDCs where neutron background events concentrate. The PDC
response in the \mbox{10 MeV} range and the validity of the Geant4
simulations of the in-flight neutron background described in
Section~\ref{Background sim} were tested with \mbox{14 MeV} neutrons
from a D--T reaction. In these tests, a simplified detector geometry,
consisting of four fast plastic scintillators and three SAS BGO
crystals as shown in Fig.~\ref{Gothenburg array}, was used. This detector array was surrounded by a \mbox{10 cm} thick polyethylene shield, mimicking the PoGOLite detector construction.
The neutron count rate in the central scintillator (Fast0) was measured both with and without active vetoing in the surrounding scintillators. 
In both cases, the recorded spectrum was well reproduced in simulations. A example is given in Fig.~\ref{SimMeas_10cm}, which shows the measured and the simulated spectrum obtained with the vetoing system in use.
The neutron count rate was reduced by \mbox{(58$\pm$4)\%} after active vetoing, and the corresponding result from simulation is \mbox{(57.0$\pm$1.4)\%}. This agreement indicates that Geant4 and the \mbox{QGSP\_BERT\_HP} physics list provide a reliable reconstruction of neutron interactions in the PoGOLite instrument. Further details of these measurements and simulations will appear in a separate publication (Kiss et al.~\cite{MK_Gothenburg}).

Simulation studies show that charged cosmic ray interactions in the
PDC assembly can be f\mbox{}iltered out to a negligible level because of
their higher energy deposition.
However, saturation effects due to large signal depositions could mimic
low-level fast scintillator signals in subsequent events.  Moreoever,
in the SAS, cosmic ray crossing can introduce dead time and lower
offline event f\mbox{}iltering eff\mbox{}iciency.  Various tests were
conducted with a proton beam (392~MeV) at the Reseach Center for Nuclear
Physics in Osaka to study the behavior of the front-end electronics under
cosmic ray bombardment. In a test run, a gamma-ray spectrum from $^{241}$Am irradiating the fast scintillator was recorded while protons bombarded
either the slow scintillator tube or the bottom BGO of the PDC. We
f\mbox{}ound that the spectrum is little affected down to about 20~keV
for a proton rate of 5~kHz, which is more than an order of magnitude
higher than the expected cosmic ray rate. The measurement described in
section 2.2 and Fig.~\ref{Sr90psd2} also show that clean fast signals
(gamma-rays from $^{241}$Am) are recovered even under an intense electron
bombardment ($\sim$10~kHz) of the slow scintillators.

The expected PoGOLite performance as given in Table~\ref{Table1} has been calculated based upon beam test results and simulations.
By reducing backgrounds and having a large effective area
(about 228~cm$^2$ at 40~keV),
PoGOLite is able to detect a 10\%
polarised signal from a 200~mCrab source
in a 6~hour f\mbox{}light.


\section{Science with PoGOLite}

\subsection{Crab Nebula}

The Crab Nebula has been the prime target of polarisation measurements
in several electromagnetic wave bands. The f\mbox{}irst detection of
large polarisation in the optical band, by Vashakidze~\cite{Vashakidze54},
Oort \& Walraven~\cite{Oort&Walraven56} and Woltjer~\cite{Woltjer57},
led to the establishment of synchrotron radiation as the dominant emission
mechanism in the optical band, similar to the process suggested
as an explanation for the polarised radio emission by
Shklovsky~\cite{Shklovsky53}.
Stimulated by the implication that electrons are accelerated
to high energies in the nebula, many X-ray observations followed, including
high spatial resolution ($\sim$15~arcsec) mapping of the nebula
in the X-ray band (1--6~keV) by Oda et al.~\cite{Oda67}.
The authors concluded that the X-ray source
is extended and that its intensity prof\mbox{}ile is consistent
with its optical image. Several high spatial resolution maps
were obtained in the soft gamma-ray band using lunar occultation
of the Crab nebula (see references given in
Aschenbach \& Brinkmann \cite{Aschenbach&Brinkmann75}).

Aschenbach \& Brinkmann~\cite{Aschenbach&Brinkmann75} modeled the
X-ray emitting structure of the Crab nebula based on observations available then and concluded that high energy electrons are trapped in a
magnetic torus around the Crab pulsar and emitting synchrotron
radiation. We note that Rees \& Gunn~\cite{Rees&Gunn74} demonstrated
theoretically that a magnetic torus will be built up by the
relativistic outf\mbox{}low of the Crab pulsar.  The plane
def\mbox{}ined by the magnetic torus was predicted to be perpendicular
to the spin axis of the Crab pulsar.  The radiation region can extend
well beyond the torus and its size is determined by the magnetic
f\mbox{}ield and involves both electron diffusion and bulk motion.
The diameter was calculated to be around 1~arcmin at 20~keV and
\mbox{2--4 arcmin} in the optical band by Aschenbach \& Brinkmann~\cite{Aschenbach&Brinkmann75}.
The authors noted that the observed
elongation in the optical emission region was orthogonal to that in the X-ray emission region, probably due to propagation conditions of the
electrons. For higher energy X-rays, the emission region was predicted
to shrink to an extended elliptical conf\mbox{}iguration with a major
axis of about 40~arcsec and a minor axis of 22~arcsec.

The entire Crab nebula was imaged in the soft gamma-ray band
(22--64~keV) to about 15~arcsec
resolution by Makishima et al.~\cite{Makishima81} and
Pelling et al.~\cite{Pelling87}.
The observed images were consistent with the emission being
synchrotron radiation from high energy electrons trapped around
a magnetic torus similar to that described above. The radiation
region in the 43--64~keV band was signif\mbox{}icantly smaller
than in the 22--43~keV band, consistent with
the prediction by Aschenbach \& Brinkmann~\cite{Aschenbach&Brinkmann75}.

Fine X-ray images taken with the Chandra X-ray Observatory have brought a renaissance
to the structural study of the Crab nebula and other pulsar wind nebulae
(see Weisskopf et al. \cite{Weisskopf00} and references given in
a review by Arons \cite{Arons04}). The discovery of two concentric
torii is one of the many key observations brought forward by Chandra.
Ng \& Romani~\cite{Ng&Romani04} used the Chandra image to
determine the orientation of the magnetic torii accurately:
the position angle of the Crab pulsar spin axis was determined
to be $124.0\pm 0.1$ degrees and
$126.31\pm 0.03$ degrees on the inner and outer torii, respectively.
The angle relative to the line of sight was determined
to be $61.3\pm 0.1$ degrees and
$63.03^{+0.02}_{-0.03}$ degrees for the inner and outer torii, respectively.

If high energy electrons are trapped in a macroscopic toroidal magnetic
f\mbox{}ield, we expect the
polarisation plane to be parallel to the spin axis.  Nakamura \&
Shibata~\cite{Nakamura&Shibata07} built a 3D model of the Crab nebula
and predicted the polarisation position angle dependence.
A 3D relativistic MHD simulation of pulsar wind nebulae was
presented by Zanna et al.~\cite{Zanna07}.  Both models
are axisymmetric and predict the polarisation plane to be
parallel to the spin axis.
Zanna et al.~\cite{Zanna07} predicts
the surface brightness distribution be signif\mbox{}icantly different
in the optical and X-ray bands.

The X-ray polarisation position angle measured by Weisskopf et al.
\cite{Weisskopf76} is $161\pm 2.8$ degrees at 2.6~keV and $155.5\pm 6.6$
degrees at 5.2~keV, i.e. about 30~degrees from the spin axis determined
by Ng \& Romani~\cite{Ng&Romani04}.
The polarisation angle is consistent with $162.0\pm 0.8$ degrees
measured in the optical band
of the central region (diameter 30~arcsec) of the Crab nebula
(Oort and Walraven \cite{Oort&Walraven56}). A spatially resolved
polarisation measurement of the entire Crab nebula by Woltjer~\cite{Woltjer57}
has revealed that the emission is extended
to 4 arcmin by 3 arcmin and highly polarised
(up to $\sim$50\%) in the outer part of the nebula.
In the central part, the measured degree of polarisation
fluctuates highly at 10~arcsec scale, possibly associated
with filamentary structures. Woltjer~\cite{Woltjer57} has estimated the error
in his position angle measurement to be $\sim$6~degrees.

As the emission region shrinks for higher energies, the polarisation
position angle is expected to approach that of the spin axis
($\sim$124--126~degrees). PoGOLite can isolate the Crab nebula
from the Crab pulsar by removing gamma rays detected within 1.65~ms and
3.3~ms of the two peaks, P1 and P2 in Fig.~\ref{CrabPulsarModels},
respectively. PoGOLite can detect 2.7~\% polarisation (3~$\sigma$)
and determine the position angle with a precision of $\sim$3-4~degrees
for the nebula component between 25--80~keV.
Energy dependence of the polarisation
position angle will either confirm
the magnetic torus around the Crab pulsar
or seriously challenge the standard model of high energy emission. Either way, it will make a
critical contribution to the modeling of the Crab nebula.

\subsection{Emission Mechanism in Isolated Pulsars}

Rapidly-rotating neutron stars are a prime target for polarimetric studies.
Their emission region is known to
be different in different wavelength bands. Historically, phase-resolved
polarimetry has had enormous diagnostic capability at radio and optical
wavelengths. The expected signature of emission near the poles of a dipole
f\mbox{}ield, an 'S'-shaped swing of the polarisation
position angle through the
pulse prof\mbox{}ile (Radhakrishnan \& Cooke \cite{Radhakrishnan69}),
has been seen in many radio pulsars.
This is generally accepted as proof that the radio emission originates
from the open f\mbox{}ield lines of a magnetic dipole.
In the X-ray and gamma-ray
regimes, each pulse period reveals two peaks, called P1 and P2. A phase-resolved polarisation measurement
of the corresponding soft gamma-ray f\mbox{}lux will indicate
where in the magnetosphere the emission occurs.

\begin{figure} 
\centering
\includegraphics[width=8.8cm]{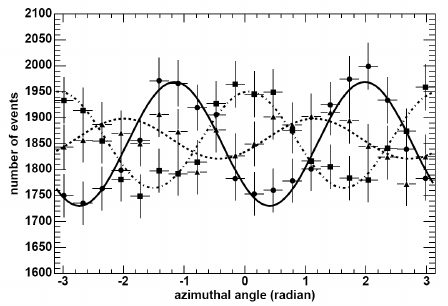}
\caption{Modulation in the azimuthal Compton scattering angle predicted
for P1 of
the Crab pulsar by 3 pulsar models: (solid) polar cap model,
(dashed) outer gap model, (dot-dash) caustic model.
A 6 hour simulated
observation by PoGOLite at 4~g/cm$^2$ atmospheric overburden is shown.}
\label{P1ModulationBW}
\end{figure}

Models for the high-energy emission fall into two general classes. In the
polar cap model and its various modif\mbox{}ications (Sturner et al.
\cite{Sturner95} and Daugherty \& Harding \cite{Daugherty96}),
particle acceleration occurs near the neutron star surface and the
high-energy emission results from curvature radiation and inverse-Compton
induced pair cascades in the presence of a strong magnetic f\mbox{}ield.
Outer gap models (Cheng et al. \cite{Cheng86}, Romani \& Yadigaroglu
\cite{Romani95} and Romani \cite{Romani96}) assume that
acceleration occurs
in vacuum gaps that develop in the outer magnetosphere along the last
open f\mbox{}ield line between the null-charge surfaces and
the light cylinder,
and that high-energy emission results from electron-positron cascades
induced by photon-photon pair-production.
These mechanisms intrinsically produce highly
polarised radiation (up to 70\%, depending on the particle spectrum)
beamed along the magnetic f\mbox{}ield lines, with electric vectors parallel
or perpendicular to the local f\mbox{}ield direction.
\begin{figure*}[t] 
\centering
\includegraphics[width=18cm]{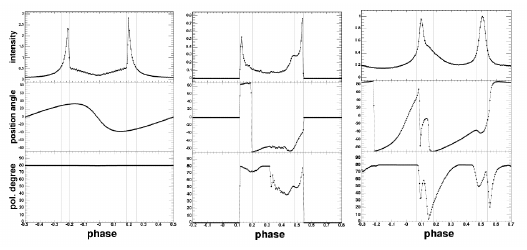}
\caption{Model predictions versus phase: intensity (top panels),
polarisation position angle (middle panels)
and polarisation degree (bottom panels) for (left) the polar cap model
(Daugherty \& Harding \cite{Daugherty96}),
(center) the outer gap model (Romani \& Yadigaroglu \cite{Romani95}) and
(right) the caustic model (Dyks \& Rudak \cite{Dyks03}).
The areas between a pair of vertical lines in the figure correspond
to the first pulse (P1) and the second pulse (P2).}
\label{CrabPulsarModels}
\end{figure*}
The Crab pulsar is the primary target for PoGOLite. Fig.~\ref{CrabPulsarModels}
shows the theoretical polarisation position angle and polarisation degree as
a function of pulse phase for the polar cap, outer gap and slot gap or
`caustic' (Dyks \& Rudak \cite{Dyks03} and Dyks et al. \cite{Dyks04})
models. In the caustic model, the Crab pulse
prof\mbox{}ile is a combination of emission from both poles, whereas in both
the polar cap and outer gap models, radiation is seen from only one pole
or region. The signature of caustic emission is a dip in the polarisation
degree and a rapid swing of the position angle at the pulse peak.
The expected azimuthal modulation for P1, as def\mbox{}ined
in Fig.~\ref{CrabPulsarModels},
has been calculated for a 6 hour observation and is shown in
Fig.~\ref{P1ModulationBW}. In this calculation, we assumed that the
nebula component is polarised with a polarisation degree and angle as
measured by OSO--8 (Weisskopf et al. \cite{Weisskopf76}).

The PoGOLite instrument will distinguish between these three models
at the 5$\sigma$ level using P1 alone. Predictions of the polarisation characteristics at P2
for the three models are rather
similar. Measurements of the region between the two peaks will also be
important in understanding the pulsar emission mechanism in the
gamma-ray band. Even if none of these models turns out to be correct,
the PoGOLite data will provide strong
constraints upon any future models.

\subsection{Accereting Black Holes}

Another class of intense galactic X-ray sources where soft gamma-ray
polarisation is expected are X-ray binary (XRB) systems, such as
Cygnus X--1, where a compact object accretes matter from a
companion star (for reviews, see Tanaka \& Lewin \cite{Tanaka95};
Charles \cite{Charles98}).
Dramatic spectral differences
suggest that the two major states seen in such sources are related to changes
in the mass accretion rate, resulting in different structures of
the accretion flow (see e.g. Zdziarski \cite{Zdziarski04}).
At high accretion rates, accretion takes place through a geometrically
thin, optically thick accretion disc, extending into the innermost stable
orbit around the black hole (Shakura \& Sunyaev \cite{Shakura73}).
At lower accretion rates, the inner part
of the disc is replaced by a geometrically thick, optically thin, hot inner
flow, possibly advection-dominated (Esin et al. \cite{Esin97}).
Most XRBs exhibit transitions between these two spectral states, commonly
referred to as the soft and the hard state, respectively.

In the hard state, the primary X-ray emission arises via
repeated Compton upscattering of thermal X-ray photons from the truncated
outer disc by hot electrons in the inner flow. The resulting hard X-ray
spectrum also shows signatures of a reprocessed component, arising from
the primary hard X-ray photons reflecting off the cooler accretion disc.
Evidence for this reflection component is provided by a fluorescent Fe-line at
$\sim$6.4~keV, an Fe absoption edge at 7 keV, and a broad ``hump'' at
$\sim$10--200~keV (Lightman \& White \cite{Lightman88}). While the
primary emission is
not expected to be significantly polarised as it arises via multiple
Compton scatterings, the reflection component is expected to display
significant intrinsic polarisation (Matt \cite{Matt93};
Poutanen et al. \cite{Poutanen96}). The observed degree of polarisation
is also dependent on the inclination
of the system, giving polarimetry the potential to derive
inclinations of XRB systems, which is often difficult by other means.

Some viable, but not widely accepted, models suggest that a part of the high
energy emission arises via synchrotron processes in a collimated outflow or
jet (Markoff et al. \cite{Markoff05}; see also Sect.~4.5).
These models predict a much higher degree of polarisation.

In the soft spectral state, hard X-ray emission arises from Compton
upscattering of photons by energetic non-thermal electrons distributed
in active regions above the accretion disk.  As
both the geometry and electron distribution differ from that in the hard
state, a different polarisation signature is expected. Apart from
geometrical constraints, polarisation measurements in the soft state can
provide information about such a non-thermal electron distribution.
Measuring the polarisation at energies
above 10 keV will thus provide independent constraints on the inferred
accretion geometry, and constitutes a test for the models inferred from
spectral and temporal studies.

A natural candidate for polarisation measurements is the persistent
source Cygnus~X--1, one of the brightest and best studied XRBs, where
the compact object is a stellar-mass black hole. Since the albedo of the
reprocessed component contributes $\sim$30\%, a net polarisation of
$\sim$10\% is expected in the hard state, and a slightly lower degree in the soft state.
Simulations show that even a low degree of polarisation (3--5\%) is
detectable with PoGOLite in either spectral state. In a long-duration
flight, PoGOLite will also be able to measure energy dependence of the
polarisation (see Axelsson et al. \cite{Axelsson07},
Engdeg\aa rd \cite{Engdegard06}).

The degree of polarisation may be even higher in other XRB sources, e.g.,
Cygnus~X--3. Recent results show that it may exhibit a state totally
dominated by Compton reflection (Hjalmarsdotter et al.
\cite{Hjalmarsdotter07}).

The spectral formation process
in active galactic nuclei -- mainly in Seyfert nuclei -- is believed to be
similar to that inferred for Galactic XRBs, although the
mass of the accreting black hole is at least $10^6$ times higher.
While a short PoGOLite flight is unlikely to be sufficiently sensitive to measure soft
gamma-ray polarisation in such sources (e.g. for NGC 4151, the emission is
much fainter, reaching 10--80~keV fluxes of $\sim$10~mCrab),
they certainly are excellent candidates for future polarisation studies
in the soft gamma-ray band.

\subsection{Accreting Neutron Stars}

Accretion onto highly magnetized neutron stars, such as Hercules X--1 and
4U0115+63, provides a unique opportunity to study
physical processes under extreme conditions.
The observed X-ray and gamma-ray radiation originates
from accreted material f\mbox{}lowing along
magnetic f\mbox{}ield lines onto distinct regions near the polar caps
(for a review, see Harding \cite{Harding94}). The localized emission
region and
neutron star rotation lead to a modulation of the observed flux. The timing of these pulsations can be used to determine the orbital
parameters for the neutron star in the binary system.
A harmonic absorption feature has been detected first from
Hercules X--1 by Tr\"{u}mper et al.~\cite{Truemper78}
and later from several other objects in the hard X-ray spectrum
(Coburn et al. \cite{Coburn02}).
These absorption features are interpreted as
cyclotron resonances in magnetic f\mbox{}ields of order
$10^{12}$~Gauss.

Due to the strong magnetic field, the radiation from the accretion column
is expected to be linearly polarised. As the neutron star rotates, the
orientation of the magnetic field with respect to our line of sight
changes and so does the direction and strength of the polarisation.
By observing these modulations, one can determine the orientation of
both the rotation and the magnetic axis of the star. With observations
separated in time, one can also search for neutron star
precession, which should show up most clearly in polarimetry.

The beaming pattern of the emitted radiation depends strongly on
how the accreting matter
is decelerated as it hits the neutron star. If a stand-off shock is
formed, the radiation will come from a vertically extended accretion
column with the strongest emission normal to the magnetic field (fan beam).
On the other hand, deceleration by Coulomb collisions will result in a
thin hot plasma slab with strongest emission in the vertical direction
(pencil beam), which is parallel to the magnetic field
(Meszaros et al. \cite{Meszaros88}). The two geometries will have opposite
correlations between flux and polarisation, so with polarisation
measurements, it will finally be possible to distinguish between these
two alternatives.

In both models above, the
polarisation is expected to vary with energy. PoGOLite will not
be able to study polarisation variations within a cyclotron feature
but it has sufficient energy resolution to detect a strong variation
of polarisation across the full spectral band of the instrument
(Axelsson et al. \cite{Axelsson07}).

Hercules X--1 is the highest priority target among the persistently
bright sources in this category. In addition to the persistent
sources, the galaxy also contains a population of similar transient
accreting X-ray pulsars. During outbursts,
which may last for weeks or months, some of these sources are among
the brightest X-ray sources on the sky. One example is V0332+53,
which at the end of 2004 had its fourth outburst since the early
1970s. At maximum, the source flux exceeded 1 Crab and showed 3
strong cyclotron features in its hard X-ray spectrum
(see, e.g., Mowlavi et al. \cite{Mowlavi06}). The high flux and strong
cyclotron features would make this an extremely interesting target
for hard X-ray polarimetry.

\subsection{Jets in Active Galaxies and Galactic Binaries}

Imaging and variability observations in many wavelength bands indicate that
a fraction of all active galactic nuclei are associated with relativistic
jets pointing close to our line of sight.  Emission is highly variable, and
extends over all accessible bands.  EGRET has revealed that many active
galaxies possessing jets are powerful gamma-ray emitters, and that the
observed gamma-ray f\mbox{}lux originating in the jet by far dominates that
measured in other bands (see, e.g., von Montigny et al.
\cite{vonMontigny95});  such jet-dominated active galaxies are
commonly known as blazars.  Jets are thus common and energetically important
ingredients for such classes of active galaxies.
However, the formation, acceleration,
collimation, and contents of a jet are poorly understood, although the
most promising models invoke magneto-hydrodynamic processes as the
mechanism for the jet production (see a review
by Sikora et al. \cite{Sikora05}).

From an observational standpoint, the broad-band emission spectra
of blazars generally show two pronounced humps:
one in the radio to soft-gamma-ray range, most likely produced by
the synchrotron process; and another, peaking in the MeV/GeV band and
extending in some objects up to TeV energies, most likely due to
inverse-Compton processes via the same electrons that produce
the synchrotron emission. The inference that emission from the low-energy
hump in blazars is produced by synchrotron radiation is based on
the observed high degree of polarisation in the radio through UV bands.
However, we know nothing about the level of polarisation in the X-ray/soft
gamma-ray band, although the polarisation properties are a powerful tool
to discriminate between emission models (see, e.g.,
Poutanen \cite{Poutanen94}, Sikora et al. \cite{Sikora94},
Lazzati \cite{Lazzati05}). In one sub-class of blazars,
the so-called High frequency-peaked BL Lac (HBL) objects, the synchrotron
hump def\mbox{}initely extends to the soft gamma-ray band, and the level
of polarisation as a function of the energy can be a powerful tool to study the
details of the distribution of particle energy and the intensity of magnetic
field in the jet.  PoGOLite is expected to detect high polarisation
($>$10\%) in these objects;  non-detection of polarisation,
on the other hand, would put many current models of blazars into question.

Blazars are variable, with {\it{high states}} (occurring every few years)
lasting for months.  There are two primary Northern targets for PoGOLite
balloon f\mbox{}lights, Mkn 501 (see Kataoka et al. \cite{Kataoka99}),
and 1E1959+65 (see Giebels et al. \cite{Giebels02}).
The brighter one will be selected, depending on the current
f\mbox{}lare state.
In the Southern hemisphere, the most promising target is PKS 2155--304
(see, e.g., Kataoka et al. \cite{Kataoka00}).

The Galactic analogues to quasars, X-ray binary systems, also sometimes
show powerful jets, and are then referred to as microquasars,
as reviewed by Fender et al. \cite{Fender04}. The jets in microquasars are
studied in the radio band, as the emission region is too compact to be
resolved in X-rays. A detection of a high degree of polarisation
in the soft gamma-ray band would support
the jet origin of gamma-rays, a model that is viable but not at all
widely accepted (see Fender et al. \cite{Fender04}). Since the radiation
is from high-energy
electrons trapped in jets, a measurement of the plane of polarisation
will reveal the direction of the magnetic f\mbox{}ield. The prime microquasar
target is GRS 1915+105, which is the most spectacular of such sources,
often reaching soft gamma-ray f\mbox{}luxes twice that of the Crab system (for
a recent overview, see Zdziarski \cite{Zdziarski05}).

\begin{table}
\centering
\caption{Expected PoGOLite Northern Targets}
\label{Targets}
\begin{tabular}{lrr}
\hline\hline
Target & Coincidence rate & \hspace{2mm} Min. Det. Pol. \\
\hline
Crab (total) & 15.2/s & 2.4\% \\
Cyg X-1 (Hard state) & 14.9/s & 2.4\% \\
Cyg X-1 (Soft state) & 5.3/s & 4.6\% \\
Hercules X-1 & 2.7/s & 8\% \\
Mkn 501 (Flare) & 0.82/s & 17\% \\
V0332+53 (burst) & $\sim 4/$s & 5.3\% \\
4U0115+63 (burst) & $\sim 4/$s & 5.3\% \\
GRS 1915 (burst) & $\sim 4/$s & 5.3\% \\
\hline\hline
\end{tabular}\label{Table2}
\end{table}


\section{Summary and Conclusions}

PoGOLite is a balloon-borne soft gamma-ray polarimeter designed to
enhance the signal-to-noise ratio\comment{Added by TK on 2007-10-12.}
and secure a large effective area for polarisation measurement
($\sim$228~cm$^2$ at 40~keV)
by limiting the field of view of individual pixels to
1.25~msr (FWHM)
based on well-type phoswich detector technology.
Thick neutron and gamma-ray shields have been
added to reduce background to
a level equivalent to $\sim$100~mCrab between 25--50~keV.  Through
extensive detector characterisation at polarised synchrotron beams and
tests with neutrons, radioactive sources and accelerator protons, we have
confirmed the predicted instrument to \comment{Was ``performance"}
perform in accordance with Geant4-based
simulation programs. PoGOLite can detect a 10\% polarisation from a
200~mCrab object in a 6~hour balloon flight and will open a new observational window
in high energy astrophysics.

\section{Acknowledgments}

We wish to thank Jonathan Arons, Makoto Asai, Roger Blandford, Blas Cabrera,
Pisin Chen, Persis Drell, Steven Kahn, Tatsumi Koi, Kazuo Makishima,
Takashi Ohsugi, Masashi Takata,
and Marek Sikora for their continuing support and encouragement.
We are grateful to Alice Harding for providing pulsar model
predictions in numerical form. We acknowledge
Charles Hurlbut and Eljen Technology for developing the
slow scintillator tube, Hamamatsu Photonics for improving the
performance of R7899EGKNP, and
the Nikolaev Institute of Inorganic Chemistry for
supplying BGO crystals of excellent quality.
T.K. has benef\mbox{}itted greatly
from discussions with Jonathan Arons on emission mechanisms in
pulsar wind nebulae.
During the early developmental phase, the PoGOLite Collaboration
benefitted from discussions with
John Mitchell, Robert Streitmatter and Daniel Marlow.

We gratefully acknowledge support from the Knut and Alice Wallenberg
Foundation, the Swedish National Space Board, the Swedish Research Council,
the G\"{o}ran Gustafsson Foundation,
the U.S. Department of Energy contract to SLAC no. DE--AC3--76SF00515,
the Kavli Institute for Particle Astrophysics and Cosmology (KIPAC)
at Stanford University through an Enterprise Fund, and the Ministry of
Education, Science, Sports and Culture (Japan) Grant-in-Aid in Science
No. 18340052. J.K. and N.K. acknowledge support by JSPS Kakenhi No. 16340055.
J.K. was also supported by a grant for the international mission research
provided by the Institute for Space and Astronautical Science (ISAS/JAXA).
T.M. acknowledges support by Grants-in-Aid for Young Scientists (B) from
JSPS (No. 18740154).

\bibliographystyle{aa} 

\end{document}